\documentclass{article}

\usepackage{arxiv}

\usepackage[utf8]{inputenc} 
\usepackage[T1]{fontenc}    
\usepackage{hyperref}       
\usepackage{url}            
\usepackage{booktabs}       
\usepackage{amsfonts}       
\usepackage{nicefrac}       
\usepackage{microtype}      
\usepackage{lipsum}		
\usepackage{graphicx}
\graphicspath{{images/}}
\usepackage{natbib}
\usepackage{doi}

\usepackage{siunitx}
\DeclareSIUnit\angstrom{\text {Å}}
\usepackage{tabularx}
\usepackage{authblk}
\usepackage{makecell}

\title{Boosting Heterogeneous Catalyst Discovery by Structurally Constrained Deep Learning Models}

\date{June 2, 2022}	

\title{Boosting Heterogeneous Catalyst Discovery by Structurally Constrained Deep Learning Models}

\author[1, *]{Alexey N. Korovin}
\author[1]{Innokentiy S. Humonen}
\author[1]{Artem I. Samtsevich}
\author[1]{Roman A. Eremin}
\author[1]{Artem I. Vasilyev}
\author[1]{Vladimir D. Lazarev}
\author[1]{Semen A. Budennyy}

\affil[1]{Artificial Intelligence Research Institute (AIRI)}
\affil[*]{Corresponding author: Alexey N. Korovin, korovin@airi.net}

\renewcommand{\shorttitle}{\small{Boosting Heterogeneous Catalyst Discovery by Structurally Constrained Deep Learning Models}}

\begin{document}
\maketitle

\begin{abstract}
The discovery of new catalysts is one of the significant topics of computational chemistry as it has the potential to accelerate the adoption of renewable energy sources. Recently developed deep learning approaches such as graph neural networks (GNNs) open new opportunity to significantly extend scope for modelling novel high-performance catalysts. Nevertheless, the graph representation of particular crystal structure is not a straightforward task due to the ambiguous connectivity schemes and numerous embeddings of nodes and edges.
Here we present embedding improvement for GNN that has been modified by Voronoi tesselation and is able to predict the energy of catalytic systems within Open Catalyst Project dataset. 
Enrichment of the graph was calculated via Voronoi tessellation and the corresponding contact solid angles and types (direct/indirect) were considered as edges’ features and Voronoi volumes were used as node characteristics. The auxiliary approach was enriching node representation by intrinsic atomic properties (electronegativity, period and group position).

Proposed modifications allowed us to improve the mean absolute error of the original model and the final error equals to \SI{651}{\milli\electronvolt/atom} on the Open Catalyst Project dataset and \SI{6}{\milli\electronvolt/atom} on the intermetallics dataset.
Also, by consideration of additional dataset, we show that a sensible choice of data can decrease the error to values above physically-based \SI{20}{\milli\electronvolt/atom} threshold.
\end{abstract}


\section{Introduction}
Catalysis is a key and emergent concept in renewable energy enabling high yield, strongly specific chemical processes. This concept is in high demand in the field of chemical engineering, renewable energy and batteries. An general target of research is to develop more efficient catalysts, which stands for quite different aspects. Among others, this can be a higher activity, higher selectivity, longer lifetimes, better availability or preferable materials.\citep{Seh_2017_Theory_and_exp,kulkarni2018catalytic_trends} 
Despite remarkable achievements in materials design and, particularly, heterogeneous catalysts development for various chemical processes such as water splitting and carbon dioxide reduction, etc., the experimental trial-and-error approach is still predominate. There are difficulties associated with transition from empirical, experimental approaches towards predictive catalyst design. The main ones lie in the fact that ordinary heterogeneous catalysis process time span and spatial scale is of higher orders of magnitude than \textit{ab initio} computational chemistry can cover.\citep{ulissi2017address} Moreover, catalyst performance depends on many variables, such as the catalyst composition, morphology, support material, and reaction environment (for example, temperature, solvent, and external potential).

Development of the computational chemistry approach providing an understanding of individual elementary processes of catalytic reactions is the key to overcome the time span and spatial scale problem. However, such approach significantly reduces the complexity of the problem and considers single-crystal of actual catalyst material \citep{ertl2010reactions}. Modern computational methods for the investigation of catalytic processes mainly includes: a) density functional theory (DFT) or high-throughput theoretical calculations to estimate the energetic parameters of the elementary steps in catalysis reactions \citep{artrith2014understanding,hansen2008surface}; b) reactive force field (ReaxFF)-based molecular dynamics simulations -- prediction of real-time reaction dynamics \citep{boes2016neural,senftle2017methane}; c) kinetic (e.g. mean-field or Monte Carlo) modeling, which can simulate reaction dynamics at the time scales where the experimental reaction conditions often take place.\citep{kimaev2020artificial,peters2017reaction,yu2020carbon}
Still, such a simplification for model catalyst is a good starting point for the more efficient data-driven search for a new catalyst material with potentially outstanding properties.
However, presented classes of \textit{ab initio} methods are still too computationally expensive to cover enormously large region of interest in the configurational and compositional spaces and are confined to the only specific spatial and time scales at which those methods were developed.

Machine learning (ML) is a robust tool that can be used to learn the relationship between the input features of a assumed catalyst and its predicted output performance.\citep{Li_Wang_Xin_2018_AI_in_Cat, kitchin2018ml_in_cat} Due to the ability to learn from raw data, deep learning (DL) approaches have boosted the research and made it possible to discover new catalyst structures and extract the underlying correlated features for catalytic reactions \citep{artrith2014understanding,jinnouchi2017predicting,ulissi2017address,yang2019machine, Li_Zhang_Liu_2017ANN_cat}. For example, DL has been used to predict the catalytic behavior of materials in the competition named Open Catalyst Project (OCP) \citep{OC20,OC20_electrocatalyst}, which is driven by the DL methods materials design. 

Below, we discuss the pros and cons of graph neural networks (GNN) models to the OCP challenge and compare our solution with other models. Existing the solutions still cannot accurately  approximate  the DFT calculation results within the OCP dataset. We also speculate about the origin of this. Finally, we show how GNN-based model provided within the OCP challenge is able to predict energetics of intermetallics, which is a more specific task providing two orders of magnitude smaller error of predictions.
\section{Computational methods}
\subsection{Models}
Graph neural networks (GNN) are able to solve a wide range of problems/tasks related to data in graph format, including the prediction of a numeric property of an entire graph.
In turn, molecules and crystal structures can be considered as graphs in a natural way, which allows one to apply GNN for energy and force prediction tasks.\citep{bronstein2021geometricDL} 
Nevertheless, GNNs have some principal restrictions: they operate with a graph as a formal object  (a set of nodes and edges), while chemical compounds are complex spatial systems and thus, two isomorphic graphs can represent completely different structures.\citep{ishiguro2020weisfeiler} 
To overcome this drawback, modern models utilize information about the relative atoms’ positions as nodes’ and edges’ features and use invariant (and/or equivariant) convolution operators.

There are developed energy-rotation-invariant (and force-rotation-equivariant) GNN-based models of different architecture, size, complexity and performance (\textit{CGCNN} \citep{CGCCN}, \textit{SchNet} \citep{SchNet}, \textit{DimeNet} \citep{DimeNet} and \textit{GemNet} \citep{GemNet}, etc.). Trying to achieve a trade-off  between speed, quality and flexibility of the model's architecture (ease of encoding of angular information about node’s neighbors), we chose the \textit{SpinConv} \citep{SpinConv} model. 

All models mentioned above are capable to encode mutual atomic arrangements \textit{via} edge-to-edge angles, dihedral angles, spherical projections, etc. and various spatial convolution operators. \textit{SpinConv} uses Gaussian smearing and trainable linear transformation for distance encoding. Additionally, projection onto the unit sphere with 1D convolution in the longitudinal direction of local spherical coordinate systems is applied for aggregating information about node neighbours to embedding. These distinction determine such physical restrictions as a rotation equivariance and a difference in radii of atoms.

However, except for \textit{CGCNN}, all models (including \textit{SpinConv}) do not consider chemical information for atoms, bounding themselves only with atomic numbers (or providing convolutional layers to learn all this information). 
Also, all models turn chemical compounds into graphs using a radius-graph,
which, as presented before, seems non-physical and compels the convolution operators to handle excess neighbours.

\subsection{Crystal graph modifications}
\subsubsection{Graph connectivity and edge properties}

As mentioned, the most commonly used approach (also implemented in \textit{SpinConv}) to set graph connectivity is the radius graph, which considers atoms located closer than the predefined cutoff $R_{cut}$ as connected, taking into account periodic boundary conditions (PBCs). Thus, the intrinsic symmetry properties of a crystal structure as well as anisotropic nature of crystalline materials are vanished by the reconstruction of graph connectivity. Furthermore, such an approach is not free of parameters, selection of which may influence the results obtained.

\begin{figure}[ht]
\center{\includegraphics[scale=0.4]{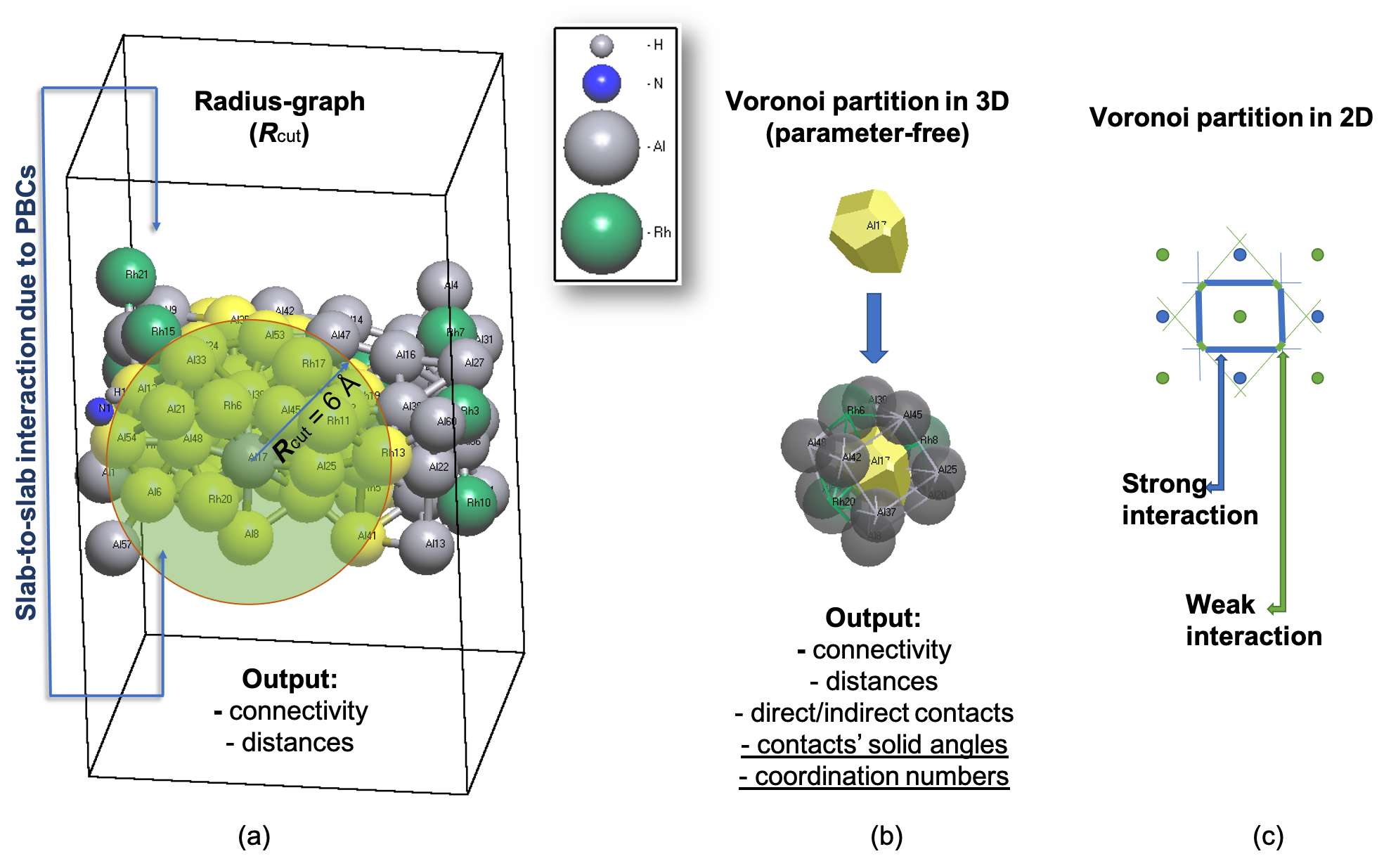}}
\caption{(a) Sample crystal structure from those provided within the Open Catalyst dataset and representation of neighbours of a certain node (atom) obtained using the radius graph method. $R_{cut} = \SI{6}{\angstrom}$. (b) For the same node of the graph, the Voronoi polyhedron and the corresponding nearest neighbours determining its shape. (c) 2D example of Voronoi partition demonstrating connection of interaction intensity differences in terms of Voronoi polygon edges.}
\label{im: voronoi}
\end{figure}

Inspired by the recent success of the CGCNN modifications\citep{park2020developing} and presence of particular GNN architectures \citep{igashov2021vorocnn}, we have considered the Voronoi partition \citep{blatov2004voronoi} of crystal space in order to modify the graph connectivity of the identical \textit{SpinConv} model. This approach is based on the real space tessellation allowing to set a number of edges of the crystal structure graph concerning only the nearest neighbours of a particular node (atom) and, subsequently, accounting for crystal symmetry in a parameter-free manner. Particularly, for the considered task, one can obtain a list of the crystal surface atoms that are most ‘open’ to the molecule. Comparison of the radius-graph connectivity and Voronoi partition is given in Figure \ref{im: voronoi}.

By using Voronoi partition instead of the crystal graph method (providing a set of distances only), one can obtain additional properties for each edge and each node of the graph (see Figure \ref{im: voronoi}b). For instance, the edge obtained for the updated connectivity can belong to direct or indirect contacts of the Voronoi net. For the formerly mentioned group, the line corresponding to the contact between the central atom and its neighbour cross the edge of the Voronoi polyhedron. This is not the case for indirect contacts. 

Additionally, the number of the neighbours (faces of the corresponding Voronoi polyhedron) can be interpreted as a coordination number of particular node of the graph (atom). For each of the neighbours (contacts), the solid angle provides information on the interaction intensity (see Figure \ref{im: voronoi}c). The higher the solid angle, the stronger the interaction between the central atom and particular neighbour. Thus, the solid angles become an additional property of the edges (contacts) of the crystal graph.

Expectedly, by using Voronoi partition we have met a problem of slab-to-slab connections through the vacuum region (see Figure \ref{im: voronoi}a). Such connections possess nonphysically large distances (anisotropic Voronoi polyhedra of large volumes), although they satisfy Voronoi tessellation rules. For this reason, additional limiting parameter $R_{lim}$ was introduced to eliminate inter-slab connections due to the periodic boundary conditions. This limiting parameter $R_{lim}$ determines the longest edge distance within the crystal graph and, in this sense, is similar to $R_{cut}$ within the radius graph approach.

\subsubsection{Node (atom) properties}
Based on the Voronoi partition, each node of the graph can be characterised by the volume of Voronoi polyhedron that represents the volume of the crystal structure ‘occupied’ by the corresponding atom. The identical \textit{SpinConv} model describes each atomic species as one-hot-encoded atomic number, which can potentially reduce generalization ability of the trained model. For this reason, we have additionally checked a number of atomic (nodal) features. 

Instead of atomic numbers, we have also considered the embedding of chemical properties previously used in CGCNN model \citep{CGCCN} including group and period numbers, electronegativity, covalent radius, valence electrons, ionization energy. Additionally, electronegativity representing an intrinsic feature of particular chemical element is a numerical type property was checked. Within an individual model, the combination of period and group (one-hot-encoded) was used instead of atomic numbers of original \textit{SpinConv}.

\subsubsection{Graph modifications}
The modifications tested within the current research are listed in Table \ref{table: models}, where RG and VG stands for the radius-graph and Voronoi graph, respectively.

\begin{table}[ht]
    \centering
    \caption{Detailed description of implemented models modifications}
    \begin{center}
    \begin{tabular}{|c|c|c|c|c|c|}
    \hline
        \textbf{Model}
        & \textbf{Short names}
        & \makecell{\textbf{Graph} \\ \textbf{connectivity}}
        & \multicolumn{2}{c|}{\textbf{Node properties}} 
        & \textbf{Edge properties}
        \\ 
        \cline{4-5}
        ~ & ~ & ~ & immanent & system-dependent & ~ 
        \\ \hline
        \makecell{VG$_{\SI{6}{\angstrom}}$ and \\ Voronoi volumes}
        & Vg+vol & VG & atomic numbers & Voronoi volumes & distances \\ \hline
        \makecell{VG$_{\SI{6}{\angstrom}}$ and \\ solid angles}
         & Vg+angle & VG & atomic numbers & - & \makecell{distances, \\ solid angles} \\ \hline
        VG$_{\SI{6}{\angstrom}}$ & Vg & VG & atomic numbers & - & distances \\ \hline
         \makecell{VG$_{\SI{6}{\angstrom}}$ and \\ electronegativity}
        & Rg+e-neg & VG & 
        electronegativity &
        - & distances \\ \hline
         \makecell{VG$_{\SI{6}{\angstrom}}$ and \\ CGCNN embeds}
          & RG+embeds
          & RG & CGCNN embeds & - & distances \\ \hline
          \makecell{VG$_{\SI{6}{\angstrom}}$ and \\ periods+groups}
         & Rg+table & RG & periods and groups & - & distances \\ \hline
        RG$_{\SI{6}{\angstrom}}$ & Rg & RG & atomic numbers & - & distances \\ \hline
    \end{tabular}
    \end{center}
    \label{table: models}
\end{table}

\subsection{Datasets}
\subsubsection{Open Catalyst Project}

Predictions of catalysts are even more challenging as combinations of small molecules and multiple crystalline surfaces should be considered.
Open Catalyst Challenge \citep{OC20}, one of recent and aspiring attempts to tackle this problem, is an open dataset combining about half a million DFT models stable and small organic reactants (up to 225 atoms).

\subsubsection{Additional dataset. Sc-Pd Intermetallics}

In order to additionally test the proposed modifications of the graph representation and to investigate the dependence of the GNN-based solution performance on the provided data, we used the intermetallics dataset constructed recently within a combined DFT/data-driven approach.\citep{intermetallics} 

\begin{figure}[ht]
\center{\includegraphics[scale=0.4]{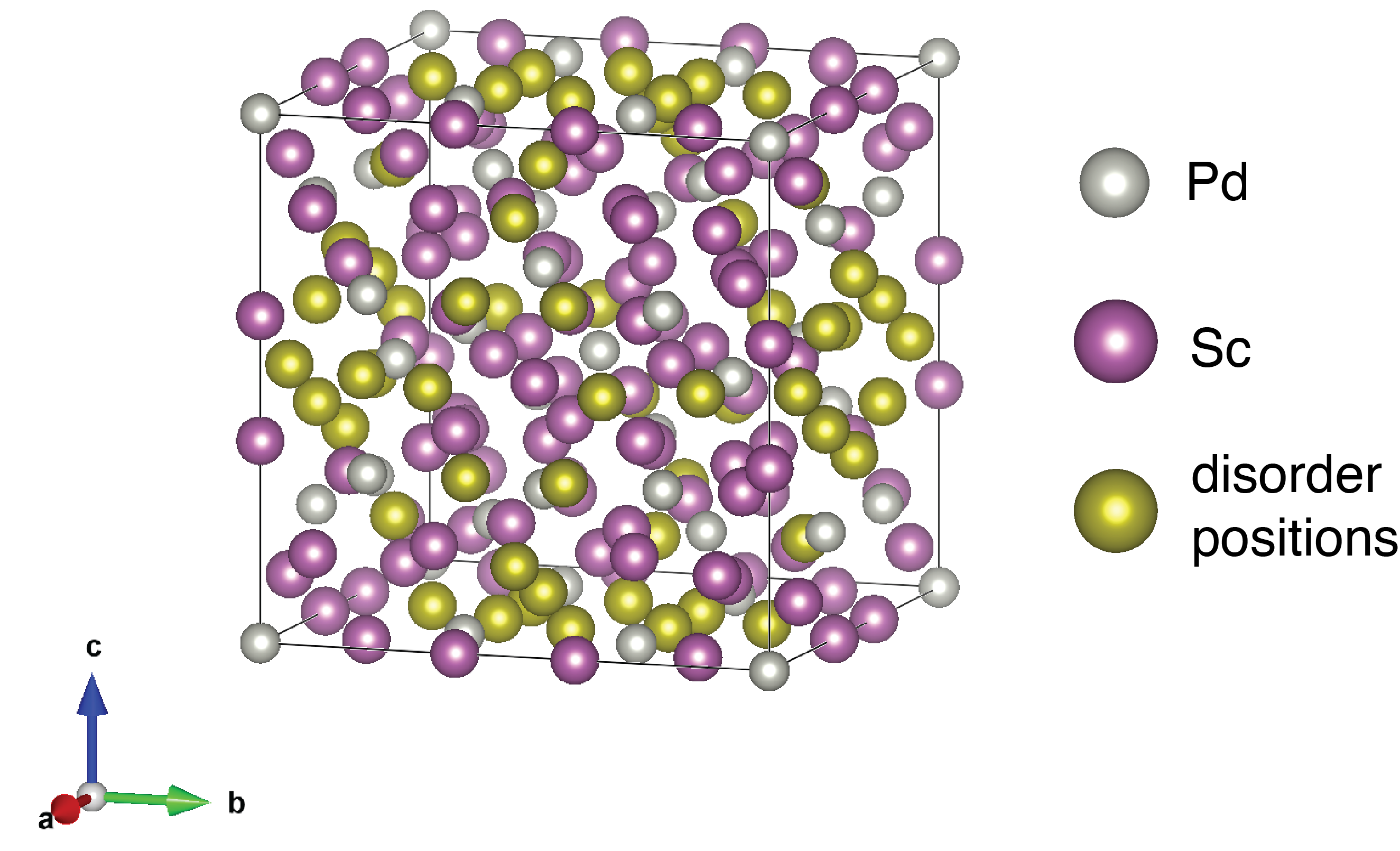}}
\caption{Sample crystal structure from those provided within the additional dataset comprising Sc-Pd intermetallics. The selected (in yellow) atomic positions correspond to the disordered region variable within the dataset.}
\label{im: intermet_fig}
\end{figure}

This data collection represents the composition/configuration space (model crystal structures) for searching for 1/1 Mackay-type \citep{tamura1997concept,Akhmetshina2017} quasicrystal approximants. 
The obtained results for the OCP dataset for structures that contain Pd show the lowest average mean absolute error (MAE) among the transitional metals (Pd, Pt, Ir, and Rh) studied within the aforementioned work (see \citep{intermetallics}).  
Each entry of the additional dataset represents a 3-periodic complex crystal structure with more than 100 metal atoms per unit cell and disordering within the region shown in Figure \ref{im: intermet_fig}. In contrast to the OCP dataset, the full DFT-based relaxation of such systems does not require restrictions neither for atomic positions (selective dynamics) nor for unit cell size/shape.

The selected Sc positions of the crystal structure in Figure \ref{im: intermet_fig} correspond to the disordering region between Mackay clusters, where vacancies and substitutions are added in different amounts during dataset generation and determining its structural and compositional diversity. Among the dataset comprising 2107 crystal structure realizations with different Sc/Pd content, among which 1896 structures (\textit{ca.} 90\%) were used within the GNN training, whereas the rest of the structures were considered for validation (211 entries, \textit{ca.} 10\%). The test dataset comprising 63 energetically favored realizations of the studied crystal structures was additionally prepared and used.

We considered DFT-relaxed energies of intermetallic crystals and the corresponding energies above the convex hull of the studied Sc-Pd system. The latter one is a good example of thermodynamic property that is less independent of the calculation scheme. For more information on the dataset preparation and content, the readers are referred to the original work.\citep{intermetallics}

\section{Results and Discussion}

\subsection{Model performances}
\subsubsection{Open Catalyst Project dataset}

For the OCP dataset, the results listed in Table \ref{table: results} clearly show influence of some of the proposed modifications of the graph connectivity and additional properties of nodes and edges (available from the Voronoi partitions) on the validation scores. It is worth noting that the proposed modifications were tested on the \textit{val\_ood\_both} datasets comprising structures with substrates and molecular fragments out of the training set domain. The results listed in Table \ref{table: results} correspond to the models trained on the full dataset (\textit{ca}. 460k structures) and 10k structure subset; the Vg+vol model is omitted in Table \ref{table: results} and excluded from the further consideration, because of its poor performance on the OCP(10k) dataset (MAE \SI{1.1}{\electronvolt/atom}). 

\begin{table}[ht]
    \centering
    \caption{For the studied datasets, the obtained MAE scores with respect to the modification of crystal graph connectivity and the features of its nodes and edges. See Table \ref{table: models} for detailed information about the models.}
    \begin{tabular}{|c|c|c|c|c|c|c|}
    \hline
        model & 
        \makecell{all \\ OCP} &
        \makecell{10k \\ OCP} &
        \makecell{RE test \\ intermet} &
        \makecell{EH test \\ intermet} &
        \makecell{RE val \\ intermet} &
        \makecell{EH val \\ intermet}
        \\ \hline
        Vg+angle & \textbf{0.651} & \textbf{0.931} & \textbf{0.02013} & \textbf{0.00579} & \textbf{0.0182} & \textbf{0.0151} \\ \hline
        Vg & * & 1.002 & 0.0234 & 0.0194 & 0.0186 & 0.0126 \\ \hline
        Rg+e-neg & 0.7096 & 0.9488 & 0.0409 & 0.01584 & 0.0259 & 0.01902 \\ \hline
        Rg & 0.6686 & 0.954 & 0.0422 & 0.0169 & 0.0262 & 0.0193 \\ \hline
        Rg+embeds & * & 1.009 & 0.046 & 0.017 & 0.0272 & 0.0188 \\ \hline
        Rg+table & 0.6731 & 0.9348 & 0.0578 & 0.0189 & 0.0292 & 0.0196\\ \hline
    \end{tabular}
    \label{table: results}
\end{table}

\begin{figure}[ht]
\center{\includegraphics[scale=0.35]{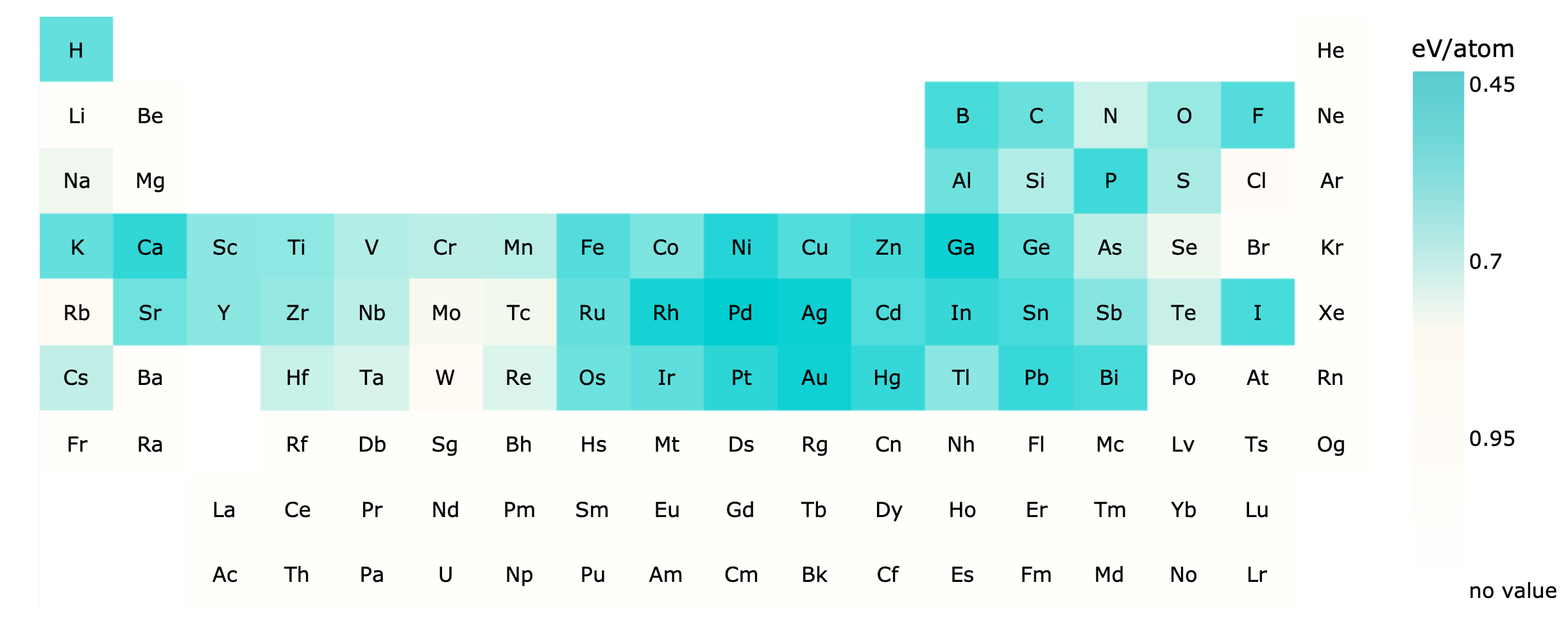}}
\caption{Characteristic mean MAE distribution averaged over the systems comprising particular chemical species within the OCP dataset for the identical \textit{SpinConv} model.}
\label{im: heatmap_table}
\end{figure}

For the OCP datasets, the best MAE value of 0.651 was obtained by using updated connectivity accompanied by the solid angles as properties of each edge of the crystal graphs of the modeled systems. Nevertheless, the difference between this improvement and the identical \textit{SpinConv} model is \textit{ca.} \SI{0.02}{\electronvolt/atom}. In turn, the additionally considered modification of embeddings of nodes (atomic species) by period/group combinations as well as accounting for their tabular electronegativities result in an increase of the corresponding MAE values of the comparable magnitudes.
An important observation at the current stage of the analysis is the fact of a non-uniform distribution of MAE with respect to the chemical compositions of the systems under consideration schematically shown in Figure \ref{im: heatmap_table}.

As one can see, the lowest MAE obtained correspond to the systems comprising transitional metals including Pt (mean MAE of 0.4953), Pd (mean MAE of 0.4317), Rh (mean MAE of 0.4628), etc. commonly used as catalysts.

\subsubsection{Sc-Pd intermetallics dataset}

The obtained results for the DFT-relaxed energies of Sc-Pd intermetallics and the corresponding energies above the convex hull for this system are given in Table \ref{table: results}. It is worth noting, that the MAE scores obtained using the identical \textit{SpinConv} model are more than 10 times smaller for the studied intermetallic compounds, which can be associated with both much poorer chemical/structural diversity resulting in a narrower target energy distribution for the studied dataset and settings of DFT-based modeling. Despite such inaccuracy of the GNN model, it was shown \citep{intermetallics} to provide reliable thermodynamic property predictions and a number of stable phases in a good agreement with available experimental data \citep{solokha2020new}.

On the other hand, the obtained MAEs are more drastically dependent on solid angles for the both studied target energies. In the case of relaxed energy, the obtained results can also be improved by the graph connectivity only (see Table \ref{table: results}). For the studied Sc-Pd systems, the lowest obtained MAE of \textit{ca.} \SI{6}{\milli\electronvolt/atom} for the energies above the convex hull is comparable to the metrics obtained by the modifications of \textit{SpinConv} architecture implemented and tested recently \citep{intermetallics}. 

\begin{figure}[ht]
\center{\includegraphics[scale=0.65]{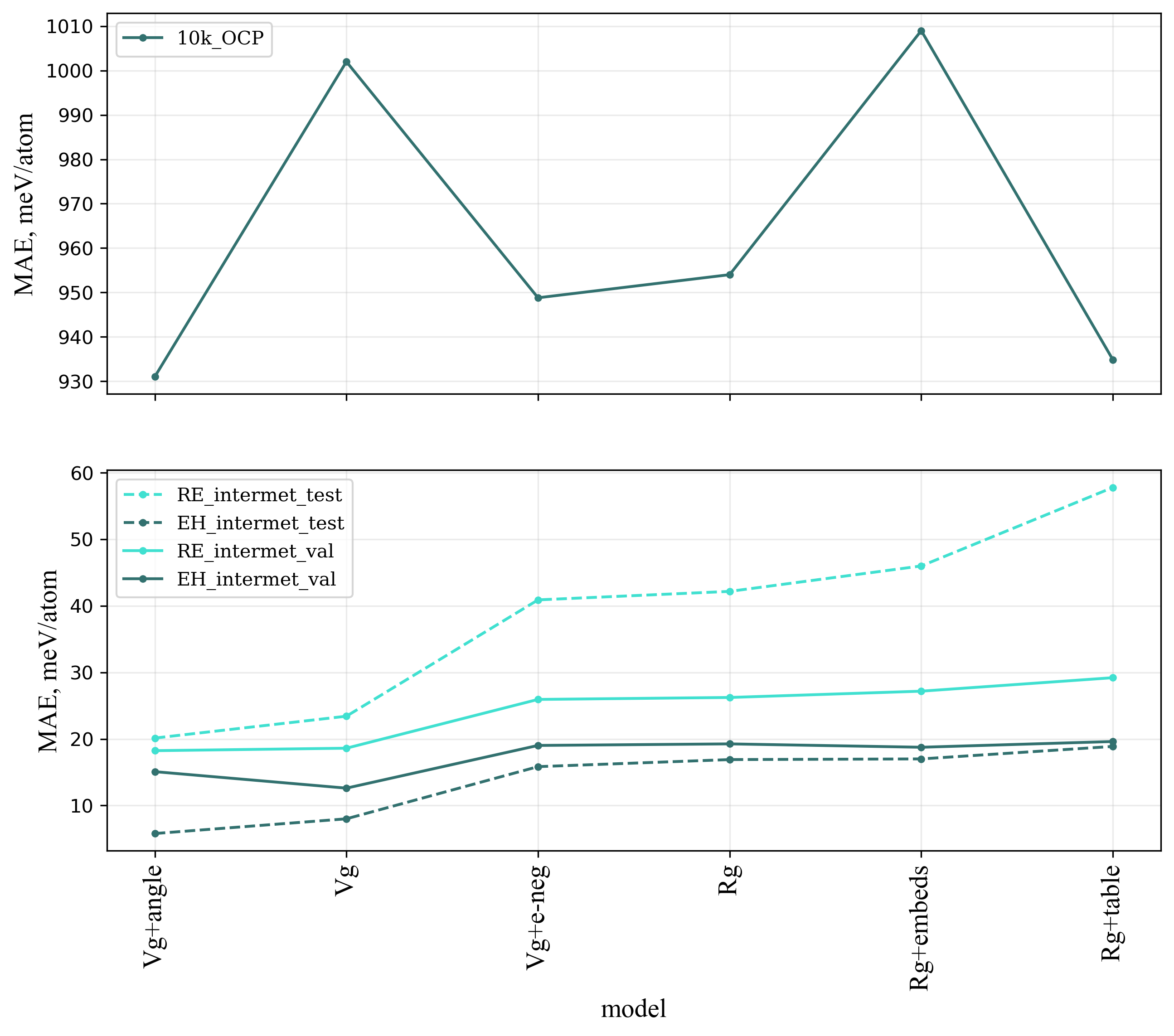}}
\caption{Performance of the models with various modifications on different samples. Top panel: performance on the OCP \textit{val\_ood\_both} dataset (models were trained on the OCP train 10k subset). Bottom panel: performance on the Sc-Pd intermetallics  validation and test datasets (models were trained on the relaxed energy (RE) and energy above the convex hull (EH) as target values). Please refer to Table \ref{table: models} for detailed information about the models.}
\label{im: results}
\end{figure}

As one can see in Figure \ref{im: results}, the implemented modifications possess stability in terms of prediction quality with respect to the subsamples (validation and test subsets) of the intermetallic dataset as well as for different target properties studied. In contrast, for the OCP dataset, the obtained results do not follow any regular pattern. This observation can be associated with the intrinsic properties of the 2-periodic (slab-like) crystal structures, for which Voronoi partition is not a straightforward (the aforementioned slab-to-slab connections) approach and requires a cutoff introduction for the graph edges.

In order to compare energy efficiency of models we measured energy consumption with help of open-source python library \textit{eco2AI}\cite{eco2ai_lib}. Carbon footprint was also estimated using regional specific carbon intensity coefficient equal to 240.56~g/kWh). As one can see from Table \ref{table: efficiency} models with precalculated Voronoi graph (Vg) were more than one order of magnitude more effective for estimation of relaxed energy (RE) comparing to models where Radius graph (Rg) was recalculated at each step. 

\begin{table}[ht]
    \centering
    \caption{For the studied datasets, the energy spent for model training and corresponding carbon emissions. See Table \ref{table: models} for detailed information about the models.}
    \begin{tabular}{ l|c|c|c|c }
    \hline
        model & \multicolumn{2}{c|}{RE} & \multicolumn{2}{c}{EH} \\ \hline
        ~ & E, kWh & $CO_2$, g & E, kWh & $CO_2$, g \\ 
        Vg+angle & 0.069 & 16.6 & 0.14 & 34.0  \\ 
        Vg & 0.069 & 16.5 & 0.25 & 59.3  \\ 
        Rg+e-neg & 1.30 & 312.6 & 1.30 & 313.4  \\ 
        Rg & 1.31 & 315.5 & 1.31 & 315.6  \\ 
        Rg+embeds & 1.32 & 317.6 & 2.22 & 320.9  \\ 
        Rg+table & 1.33 & 319.0 & 1.30 & 312.9  \\ \hline
\end{tabular}
\label{table: efficiency}
\end{table}

\subsection{Data preparation aspects}

Obviously, the presented reduction in the chemical/structural diversity of model datasets providing better test scores is limited for any high-throughput screenings or general-purpose applications. For this reason, here we want to reflect on some additional aspects of data preparation and their possible effects on the developing data-driven approaches for searching for new functional materials including heterogeneous catalysts on the basis of ‘synthetic’ (\textit{e.g.} DFT-derived) properties.

\subsection{Target properties independent on the calculation scheme}
Consideration of different target properties available from DFT calculations in connection with synthesizability of novel materials and their stability are widely discussed.\citep{bartel2020critical, peterson2021materials} Obviously, there is a direct connection between the DFT-relaxed and formation energies (or energies above the convex hull) of a particular crystal structure. Nevertheless, from computational point of view, the relaxed energies are dependent on the chosen calculation scheme (pseudopotentials, basis set size, sampling of the reciprocal space, etc.), whereas formation energies or any derivative thermodynamic properties obtained from these are free of any calculation biases. 

Moreover, the results obtained demonstrate possible differences in model prediction quality for different target energies (see Table \ref{table: results}). In turn, the descriptor-based approaches may show differences between compositional and structural feature importances, when changing target values. For instance, structural descriptors became more important features (in comparison to the chemical composition) for the energies above the convex hull (in comparison to the DFT-relaxed energies).\citep{intermetallics}

The presence of datasets that are widely diverse from chemical and structural points of view, such that of OCP, provides a great possibility for pre-training of the GNN-based models and their subsequent finetuning for the purposes of any more particular problems.\citep{intermetallics}

\subsection{Computational constrains applied}
The differences between the validation/test scores obtained for the OCP heterogeneous structures and those for the intermetallics can also be associated with computational constraints, such as selective dynamics. 

\begin{figure}[ht]
\begin{minipage}[ht]{0.49\linewidth}
\center{\includegraphics[width=0.9\linewidth]{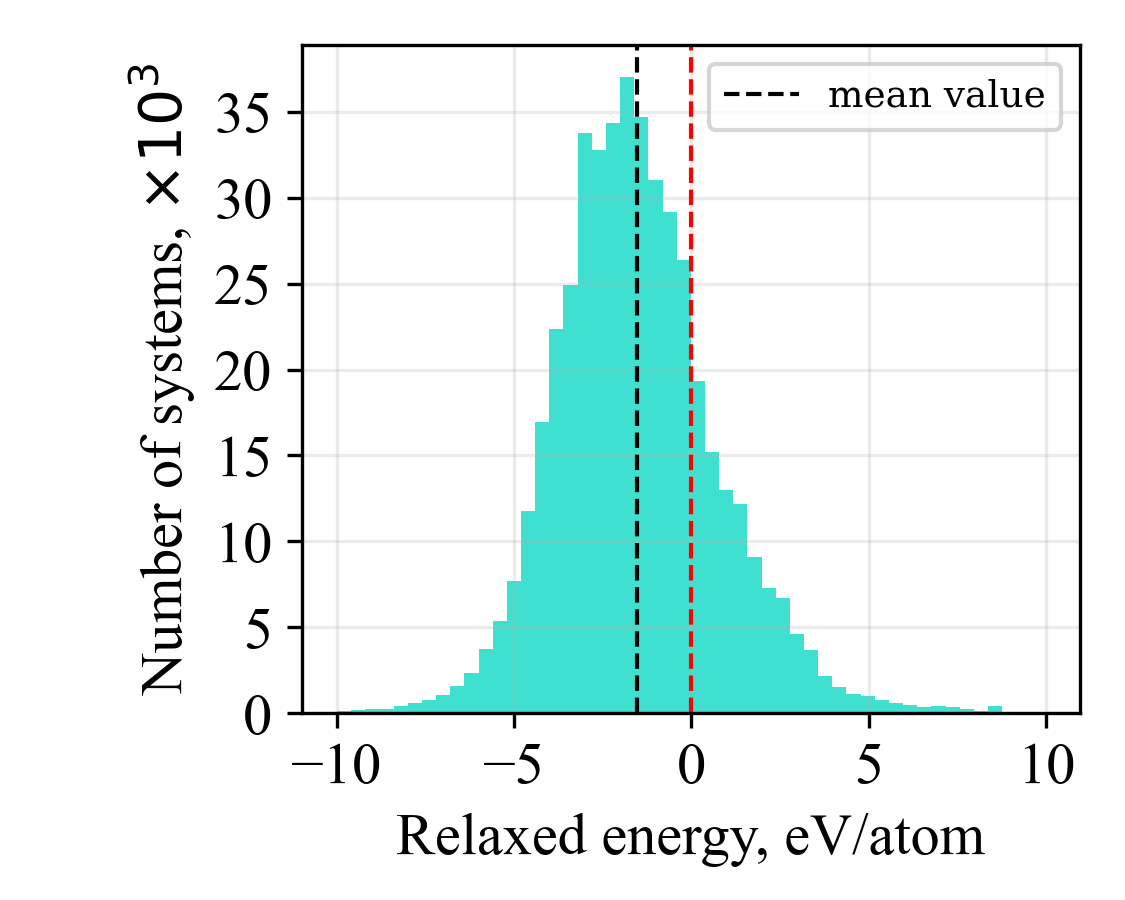} \\ a)}
\end{minipage}
\hfill
\begin{minipage}[ht]{0.49\linewidth}
\center{\includegraphics[width=0.9\linewidth]{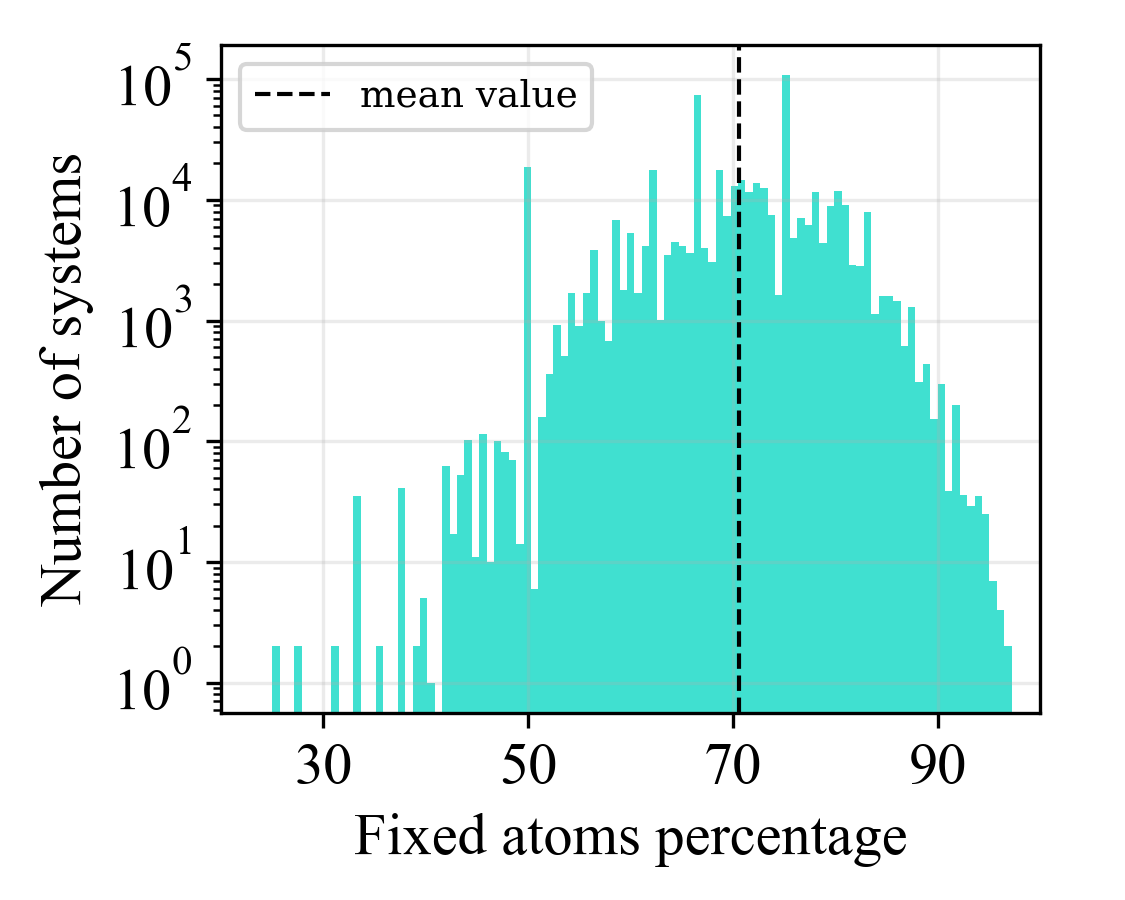} \\ b)}
\end{minipage}
\caption{(a) Distribution of target energies available for the relaxed atomic systems (relaxed energies) within the full OCP dataset, comprising ca. 460k structures; red dotted line denotes zero energy. (b) Distribution of percentage of fixed atomic positions within the slabs of the OCP structures.}
\label{im: hists}
\end{figure}

To briefly discuss this point, we present the distribution of target energies available from the OCP datasets Figure \ref{im: hists}a. For a vast range of the crystal structures, the target energies were found to be positive, which directly points at the consequences of preparing initial structure guesses comprising the crystal slab generation, placing molecular fragment, and, for sure, selective dynamics used by the authors \citep{OC20}. 

In many DFT-based approaches, the presence of systems with altered energetics (obtained within structurally constrained relaxations) is essential in many applications, such as equation of state calculations, static and dynamic stability investigations, phonon spectra calculations, etc. Such ‘negative’ examples (for instance, those with positive energies) possess also high importance for data-driven approaches. From computational perspectives, any selective dynamics can affect the way for a system going upon the energy landscape during relaxation. 

As one can see from Figure \ref{im: hists}b, percentages of the fixed atomic positions within slabs possess wide distribution. Particularly for the Initial-Structure-to-Relaxed-Energy (IS2RE) task, this  points at the variable scale of the described effect for the dataset entries. Thus, selective dynamics might provide different biases of the target properties for different systems, although each system provided within the dataset remained structurally stable during relaxation.

\section{Conclusion and Perspectives}

Within the current research, we developed and tested modifications of the graph representation using Voronoi partition in the case of the heterogeneous systems comprising combinations of crystal slabs and molecular fragments. The comparison of the results of the developed approach for the studied 2-periodic entries of the OCP dataset and 3-periodic intermetallics showed advantages of the usage of proposed graph representations, perspectives of its further improvements and customisation, and a set of the most promising features being considered further (especially, solid angles of the contacts).

Additionally, the drastic decrease in the MAE scores accompanying the change of the systems modeled from 2-periodic (the OCP dataset) to 3-periodic (intermetallics) systems was demonstrated using the same GNN-based solutions. We associate such behavior with the reduction of chemical and structural diversity for the studied dataset and avoiding computational constraints, namely, selective dynamics used for the 2-periodic systems provided within the OCP dataset.

We want to emphasise that selective dynamics itself might be applicable in the case of a preset surface and the same amount/disposition of the fixed atomic positions within this. In such a situation, the influence of the frozen degrees of freedom of the slab relaxation might have the same energetic effect for modifications of interest, such as adsorbate positioning as an initial guess, or molecular ensembles on the adsorbent surface. The described approach, for sure, can hardly be aimed at any ambitious high-throughput applications and, again, represent a routine with a reduced structural/chemical diversity of the systems being investigated.

Nevertheless, within the described modifications and through the application of data-driven solutions, it may become possible to comprehensively study a particular surface and catalytic processes of interest. In such a case, thermodynamic properties of numerous states obtained computationally allow using Boltzmann statistics to take into account both the most energetically favorable states as well as altered/metastable ones that may have lower but non-negligible frequencies of occurrence.\citep{Cheula2020, zhang2020ensembles}  From this perspective, the authors would like to mention that the implementation of the Voronoi partition instead of the conventional radius graph resulted in a reduction of the computational costs of the studied GNN-based solutions for the obvious reason of graph simplification.

\bibliographystyle{unsrtnat}
\bibliography{main}

\end{document}